\documentstyle[preprint,prd,aps,12pt]{revtex}
\preprint{
$
\begin{array}{r}
\text{hep-ph 9402328} \\ \text{EP-CPTh.A280.0194} \\ \text{LAVAL-PHY-12-93}
\end{array}
$
}
\tighten

\begin{document}
\author{B. Dion}
\address{D\'epartement de Physique, Universit\'e Laval\\ Qu\'ebec, Canada,
G1K 7P4}
\author{Luc Marleau\cite{Laval}}
\address{Centre\ de\ Physique\ Th\'eorique,\ \'Ecole\ Polytechnique\\ 91128
\ Palaiseau\ CEDEX,\ France}
\title{Skyrmion solutions by generalization \\of the Atiyah-Manton ansatz }
\date{December 1993 }
\maketitle
\draft

\begin{abstract}
We generalize the approach of Atiyah and Manton for generating Skyrmion
configurations from instantons. The result is a series whose parameters are
found directly from the chiral angle equation. The series converge rapidly
to the exact solution for a class of the Skyrme-like models (including the
Skyrme model itself) but describe with less accuracy other types of models.
We describe the procedure and discuss its advantages and limitations.
\end{abstract}
\pacs{PACS number(s): 11.30.Na, 11.30.Rd, 12.15.Cc.  }

\section{Introduction}

As a non-linear theory of pions, the Skyrme model \cite{Skyrme61} provides
an approximate description of hadronic physics in the low-energy limit. In
this theory, the nucleon emerges as a bound state of the field, or more
precisely as a soliton. The original Skyrme Lagrangian was of fourth order
in field derivatives and it reached a 30\% accuracy with respect to physical
observables. Since then, we proposed along with other authors \cite
{Marleau89-10,Marleau90-13,Jackson,Riska} a number of alternative models
which preserved the form of the original Lagrangian while extending it to
higher orders. The result was to incorporate effects due to higher-spin
mesons and thus to improve the fit of most observables.

All those models have one handicap in common however, they do not admit
exact analytic solutions. Very few soliton solutions are known as a matter
of fact, with the exclusion of the one-dimensional sine-Gordon equation ,
the KdV equation and some other special cases. This is a natural consequence
of non-linearity, a necessary constraint for a soliton-like solution to
exist.

In the absence of analytical solutions, the only alternative to numerical
treatment is the use of aptly-chosen analytical forms which provide a
sometimes sufficient approximation. Apart from greatly simplifying
calculation of physical quantities, a great deal of information can be
extracted from such an analytic form. For instance, symmetries and general
behavior of the solution are much easier to analysis, and the individual
features proper to each Skyrme-like model also become more explicit. Then,
an analytic form proves useful in the stability analysis of the soliton,
both classical and quantum, and in the calculation of multi-Skyrmion
interactions. In particular, Igarashi et al.\cite{Igarashi} have analyzed
the quantum behavior of the Skyrme model soliton on the basis of a family of
trial functions, taking account of breathing motions and spin-isospin
rotations whereas Hosaka et al.\cite{Hosaka} have examined to the two
Skyrmion interactions.

The exact form of most trial functions is largely arbitrary however, with
the sole constraint that they have the same asymptotic behavior as the
numerical solution (i.e., $F(r)=N\pi -\alpha r+{\cal O}(r^3)$ when $%
r\rightarrow 0$ and $F(r)=\beta r^{-2}+{\cal O}(r^{-4})$ when $r\rightarrow
\infty $) and that they reproduce the numerical results with a degree of
accuracy. A few years ago on the other hand, Atiyah and Manton brought
forward the idea of generating Skyrmion solutions from Yang-Mills instantons%
\cite{Atiyah}. Using this approach as a starting point, we have developed a
systematic procedure in order to parameterize any Skyrmion solution.

We first present the Atiyah-Manton idea and discuss its successes and
drawbacks. We then introduce our own systematic procedure for parameterizing
the solution and apply it directly to a few Skyrme-like models in the $N=1$
case of a single Skyrmion configuration (Section 3), and in the more general
case of $N>1$ (Section 4). We finally offer some comments on the advantages
and limitations of our method and on various ways of improving it (Section
5) for models where the chiral angle shows non-smooth behavior due to its
nonlinear nature, .

\section{The Atiyah-Manton idea}

The Atiyah-Manton idea is based on topological similarities between
Skyrmions and instantons. Before introducing it, let us first review the
Skyrme model itself. The pion fields are represented by a unitary SU(2)
matrix denoted $U({\bf x})$. In its familiar hedgehog form, it is expressed
as follows:
\begin{equation}
\label{2.1}U({\bf x})=\exp \left[ i{\bf \tau }\cdot \widehat{{\bf x}}%
F(r)\right]
\end{equation}
where F(r) is called the chiral angle or profile function of the solution.
This field configuration constitutes a map from physical space $R^3$ into
the group manifold SU(2) and is assumed to go to the trivial vacuum for
asymptotically large distances. We therefore impose $U(r\rightarrow \infty
)\rightarrow $1. From this last condition, one may derive the existence of a
topological invariant associated with the mapping. The originality of
Skyrme's idea was to identify this invariant, i.e. the winding number, with
the baryon number.

Introducing the notation $L_\mu =U^{\dagger }\partial _\mu U$, the Skyrme
Lagrangian for zero pion mass takes the form:
\begin{equation}
\label{L}{\cal L}=-\frac{F_\pi ^2}4TrL_\mu L^\mu +\frac 1{32e^2}Tr[L_\mu
,L_\nu ]^2
\end{equation}
where the first term coincides with the non-linear sigma model and the
second one acts as a stabilizer.

The Yang-Mills instanton follows from a different path. We let $A_\mu (x)$
be a gauge field and $x=({\bf x},t)$ be a vector in $R^4$. We may then
introduce the unitary SU(2) field $U(x)$ in terms of the temporal component
of $A_\mu (x)$ by the following equation:%
$$
U^{-1}(x)\partial _tU(x)=-A_t(x)
$$

This equation is solved through the method of time-ordered products by:
\begin{equation}
\label{U}U(x)=\pm {\cal T}\exp \left[ -\int_{-\infty }^tA_t(x)dt\right]
\end{equation}
with $U({\bf x},-\infty )=-1$ and $U({\bf x},+\infty )=U({\bf x})$.

The $A_t$ component of the Yang-Mills field may furthermore be expressed in
terms of the scalar superpotential $\rho $:
\begin{equation}
\label{A}A_t=\frac i2{\bf \tau }\cdot \frac{{\bf \nabla }\rho }\rho
\end{equation}

The so-called 't Hooft instantons of topological number k are then obtained
by setting:
\begin{equation}
\rho =1+\sum\limits_{i=1}^k\frac{\lambda _i^2}{(x-X_i)^2}
\end{equation}
where the $X_i$ are poles in $R^4$ and the $\lambda _i$ are positive
constants acting as scales.

The Atiyah-Manton idea then follows from identifying the field $U({\bf x})=U(%
{\bf x},+\infty )$ with the Skyrme field, with B=k. In the B=1 case of a
single soliton centered at the origin ($X_i=0$), we set:
\begin{equation}
\label{rho}\rho =1+\frac{\lambda ^2}{r^2+t^2}
\end{equation}
and find by (\ref{U}) and (\ref{A}):
\begin{equation}
F(r)=\pi \left( 1-\frac r{\sqrt{r^2+\lambda ^2}}\right)
\end{equation}

This function has the same asymptotic behavior as the numerical Skyrmion
solution:%
$$
F(r)=\left\{
\begin{array}{lc}
\pi +\alpha r+O(r^3)\text{ , } & \text{when }r\rightarrow 0 \\  &  \\
\beta r^{-2}+O(r^{-4})\text{ , } & \text{when }r\rightarrow \infty
\end{array}
\right.
$$
with $\alpha =-\pi /\lambda $ and $\beta =\pi \lambda ^2/2$. The $\lambda $
parameter, which gives a measure of the size of the soliton, is determined
by minimizing the static mass $M_S$ of the soliton. Applying this function
to the Skyrme model, we find after minimization $M_S$= 104.1 $(F_\pi /e)$,
only 0.9\% higher than the exact numerical result.

Despite this relatively impressive success however, this approach leaves
something to be desired. For one thing, the slopes at $r\rightarrow 0$ and $%
r\rightarrow \infty $ both depend on the same parameter $\lambda $, which
means that their values cannot be adjusted independently. More precisely, we
have: $\beta =\frac{\pi ^3}{2\alpha }$. Secondly, we find that the
predictive power of this simple function rapidly decreases when we consider
more elaborate Skyrme-like models such as those we present below.

Aiming to improve the fit of the exact solution while preserving the general
form of the Atiyah-Manton ansatz, we consider an extension of the series.
The following form turns out to simplify the calculations:
\begin{equation}
\label{FAM}F(r,\lambda )=\pi \left[ 1-\frac r{\sqrt{r^2+\lambda ^2}}%
\sum_{n=0}^\infty c_{2n+1}\frac{\lambda ^{2[\frac{n+1}2]}r^{2[\frac n2]}}{%
(r^2+\lambda ^2)^n}\right]
\end{equation}
where $[z]$ is the integer part of z and $c_0=1$. This solution correspond
to a general potential of the form:%
$$
\rho =\exp \left[ -\ln Q+\sum_{m=1}^\infty d_mQ^m\right] \ \text{ where, \ }%
Q=\frac{r^2+t^2}{r^2+t^2+\lambda ^2}.
$$
where for $d_m=0$, one recovers the expression in (\ref{rho}).

This new function is still characterized by a scale $\lambda $, as well as
by the coefficients $c_n$, acting as weights with respect to the corrective
terms. As we will show those coefficients are not completely free; their
value is determined once a choice has been made for the slopes $\alpha $ and
$\beta $. The scale $\lambda $ is once again determined by minimization of
the mass of the soliton.

\section{The N=1 solitons}

It proves useful to express the Lagrangian of all Skyrme-like models in the
form (for the hedgehog solution):
\begin{equation}
{\cal L}=\sum_{m=1}^\infty h_ma^{m-1}[3a+m(b-a)]
\end{equation}
where $a=r^{-2}\sin ^2F,b=F^{\prime }{}^2$ and where we have made the scale
change: $r\rightarrow \frac{eF_\pi }{\sqrt{2}}r$. The static energy is then
written as:

$$
M_S=4\pi (\frac{F_\pi }e)\int_0^\infty r^2dr\sum_{m=1}^\infty
h_ma^{m-1}[3a+m(b-a)]
$$
or as:

\begin{equation}
\label{M}M_S=4\pi (\frac{F_\pi }e)\int_0^\infty r^2dr[3\chi (a)+(b-a)\chi
^{\prime }(a)]
\end{equation}
where $\chi (x)=\sum_{m=1}^\infty h_mx^m$ and $\chi ^{\prime }(x)=\frac{%
d\chi }{dx}$. Using the same notation, the chiral equation becomes:

\begin{equation}
\label{eqc}0=\chi ^{\prime }(a)[F^{\prime \prime }+2\frac{F^{\prime }}r-2%
\frac{\sin F\cos F}{r^2}]+a\chi ^{\prime \prime }(a)[-2\frac{F^{\prime }}r%
+F^{\prime 2}\frac{\cos F}{\sin F}+\frac{\sin F\cos F}{r^2}].
\end{equation}
The Skyrme Lagrangian corresponds to the case $\chi (a)=\chi _S(a)=a+\frac 12%
a^2$.

More general models should include higher order terms \cite{Marleau90-13}.
One could consider the model for exponential or truncated geometric series:%
$$
\chi _1(a)=e^a-1=a+\frac 12a^2+\frac 1{3!}a^3+...
$$
\begin{equation}
\chi _{2,M}(a)=a-\frac 12a^2+\frac 13a^3-...-\frac 1Ma^M
\end{equation}
which corresponds to the choice $h_{m\leq M}=\frac{(-)^{m-1}}m$.

Another interesting model has been proposed by Gustaffson and Riska \cite
{Riska}which replaces the fourth order Skyrme term by a higher order term of
order $m$:%
$$
\chi _{3,M}(a)=a+\frac 1Ma^M
$$

Some other alternatives are due to Jackson et al \cite{Jackson}. They are:%
$$
\chi _4(a)=\ln (1+a)+\frac 12a^2
$$
$$
\chi _5(a)=\frac 14[1-e^{-2a}]+\frac 12a+\frac 12a^2
$$
$$
\chi _6(a)=a+\frac{a^3}{3+2a}
$$
$$
\chi _7(a)=a+\frac{a^3}{3+4a}+\frac{a^4}{1+4a^2}
$$

In this work, we shall limit ourselves to these seven models, taking $M=$20
for the $\chi _{2,M}$ model and $M=$20 for the $\chi _{3,M}$ model.

In order to fix to the coefficients in the parameterization, we use the
chiral angle equation in the $r\rightarrow 0$ and $r\rightarrow \infty $
limits. The first step of our procedure is to express the chiral angle $F(r)$
as a Taylor expansion, valid for small r:
\begin{equation}
\label{F0}F(r)=N\pi +\alpha _1r+\alpha _2r^2+\alpha _3r^3+...
\end{equation}
and to insert this expression in the chiral equation for each model. Since
the coefficients of each of the powers of $r$ on the right-hand side of the
equation must vanish individually, we find in the case of the Skyrme model
(i.e. model $\chi _S(a)$) that $\alpha _n=0$ for $n=$even and the following
relations for the lowest $n$:%
$$
\alpha _3=-\frac{\alpha _1^3}{30}\left[ \frac{4+\alpha _1^2}{1+\alpha _1^2}%
\right]
$$
\begin{equation}
\alpha _5=\frac{\alpha _1^5}{1400}\left[ \frac{40+32\alpha _1^2+11\alpha
_1^4+7\alpha _1^6}{\left( 1+\alpha _1^2\right) ^3}\right]
\end{equation}
Thus all $\alpha _i$ coefficients are determined once $\alpha _1$ is fixed.
We repeat this operation for $r\rightarrow \infty $, using:
\begin{equation}
\label{Finf}F(r)=\frac{\beta _2}{r^2}+\frac{\beta _3}{r^3}+\frac{\beta _4}{%
r^4}+...+\frac{\beta _{10}}{r^{10}}
\end{equation}
and find $\beta _n=0$ for $n=$odd and:%
$$
\beta _4=0,\ \beta _6=-\frac{\beta _2^3}{21},...
$$

The second step of the procedure is to adjust the parameters of $c_i$ to
these power series expansion using both limits:%
$$
F(r\rightarrow 0)=\pi \left[ 1-\frac{(1+c_3)}\lambda r+\frac{(\frac 12+\frac
32c_3-c_5-c_7)}{\lambda ^3}r^3-...\right] \text{ }
$$
$$
F(r\rightarrow \infty )=\pi \left[ \frac{(\frac 12-c_3-c_5)\lambda ^2}{r^2}+%
\frac{(-\frac 38+\frac 32c_3+\frac 52c_5-c_7-c_9)\lambda ^4}{r^4}+...\right]
\text{ }
$$
Equating the various powers of this expression with the $\alpha _i$ and $%
\beta _i$ coefficients of equations (\ref{F0}) and (\ref{Finf}), we find:%
$$
c_3=\frac{\alpha _1\lambda }\pi -1,\ c_5=-\frac{\beta _2}{\pi \lambda ^2}+%
\frac 12-c_3,\ c_7=-\frac{\alpha _3\lambda ^3}\pi +\frac 12+\frac{3c_3}2%
-c_5,...
$$

Thus all the free parameters of our function are fixed once we have made a
choice for the scale $\lambda $ and for the slopes $\alpha _1$ and $\beta _2$%
. As we have pointed out, the value of those three parameters is determined
by minimization of the soliton mass $M_S$ (eq.(\ref{M})).

The calculations are performed up to order $n=10$ in (\ref{FAM}). The
results we have reached are listed in Table \ref{table1}, along with the
margin of error with respect to numerical results whenever data was
available. We notice that the agreement with numerical results varies within
a wide range, depending on the model considered. The agreement is best when
the order of the Lagrangian is low (namely for the Skyrme model and the
first two Jackson et al. models), but for the other models the series does
not seem to converge very fast. This observation is related to the
appearance of a new feature in the solution as $m$ increases. Numerical as
well as analytical analysis show that the space derivative of the chiral
angle $F(r)$ begins to develop a cusp near r=1, exhibiting phase transition.

The fact that this feature appears in the $r\sim \lambda $ region signals a
limitation in our procedure. The overall shape of our analytic profile
function is largely determined by its asymptotic behavior at $r\rightarrow 0$
and r$\rightarrow \infty $, and we have no way of carrying a direct fit of
the solution in-between. When the solution is smooth (i.e., when the order
of the Lagrangian is low), this causes no difficulty; but when phase
transition occurs we cannot expect strict agreement between numerical and
analytical solutions to hold for all $r$.

Next, we consider the Atiyah-Manton approach applied to the $N>1$
spherically symmetric solution. Although they are not the lowest energy
solutions (stable solitons), the approach leads to interesting results since
time-ordering in eq. (\ref{U}) is straighforward in these cases.

\section{The N=2,3,... hedgehog solutions}

Proceeding on to the B=2 solution of the Skyrme field equation, we consider
the spherically symmetric case:
$$
\rho =1+\frac{\lambda _1^2}{(t-T_1)^2+r^2}+\frac{\lambda _2^2}{(t-T_2)^2+r^2}
$$

The minimal energy is obtained for $\lambda _1=\lambda _2=\lambda $ and $%
(T_1-T_2)$ large but finite \cite{Atiyah}. When $(T_1-T_2)\rightarrow \infty
,$ the solution has a simple form:
\begin{equation}
F_{N=2}(r)=2\pi \left( 1+\frac r{\sqrt{r^2+\lambda ^2}}\right) =2F_{N=1}(r)
\end{equation}
and its energy turns to be larger only by a minute amount ($\sim $ 0.5 ppm).
Since the introduction a parameter for a finite $(T_1-T_2)$ seems to bring
very little improvement to the solution, we shall build our $N>1$ solutions
on the solution with $(T_1-T_2)\rightarrow \infty .$

Applying this function to the Skyrme model and minimizing the mass with
respect to $\lambda $, we find $M_S=$310.8 $(F_\pi /e)$, 1.2\% higher than
the exact result $M_S=$307.1 $(F_\pi /e)$ . Similarly, if we set $%
F_{N=3}(r)=3F_{N=1}(r),$we find the minimal mass to be $M_S=$633.1 $(F_\pi
/e)$, 3.0\% higher than the exact value. We see that the results seem to
decrease in accuracy as $N$ increases.

We generalize our own approach to the $N>1$ problem by setting:%
$$
F_N(r,\lambda )=\sum_{n=1}^NF_1(r,\lambda _n)
$$

We have considered various alternatives for $\lambda _n$. For instance:%
$$
F_N(r,\lambda )=\sum_{n=1}^NF_1(r,\lambda )=NF_1(r,\lambda )
$$
$$
F_N(r,\lambda )=\sum_{n=1}^NF_1(r,n\lambda )=F_1(r,\lambda
)+...+F_1(r,N\lambda )
$$

In the $N=2$ and $N=3$ cases, the lowest mass $M_S$ is reached as $\lambda
_n=\lambda $ and $\lambda _n=n\lambda $ respectively. We show the results in
Table \ref{table2} for the $N=2$ case. Comparing the numerical and
analytical profiles we have obtained for $F(r) $, we have noticed a slight
difference. While the latter remains smooth for all r, there is a sharper
drop-off of the former near $F(r)\cong \pi $. This disagreement surely
accounts for the poorer results we obtain (see Table \ref{table2}).

\section{Advantages and limitations of the approach}

The reliability of our approach seems to depend largely on the model
considered. The conditions imposed on the $c_i$ coefficients insure a good
agreement with numerical results when $r\rightarrow 0$ and $r\rightarrow
\infty $. This agreement is conserved when the solution remains smooth for
all values of $r$, as in the case of the $N=1$ soliton in the Skyrme and
Jackson models (i.e., models in which the order of the Lagrangian is small).
But discrepancies between analytical and numerical results begin to appear
when we consider higher-order models and when the Skyrmion number $N$
exceeds 1. Smaller scale structure then appear in the intermediate region $%
(r\rightarrow \lambda )$, and the Atiyah-Manton scheme does not have any
special treatment of those cases.

In the multi-Skyrmion problem, the sharper drop-off of the profile function
near $F=N\pi $ is smoothed out. This may be caused by the fact that our
function includes a single scale, $\lambda $. We have tried the alternative
of using two different scales by setting $F_{N=2}(r,\lambda
)=F_{N=1}(r,\lambda )+F_{N=1}(r,2\lambda )$, but with little or no
improvement in that respect.

The same type of arguments go for two-phase models. Although the phase
transition of the solution is reproduced by our analytical function, this
feature is not accurately fitted. This type of model actually seems to show
chaotic behavior near $r\sim \lambda $. Numerical as well as analytical
treatment show that when we solve the differential equation for the chiral
angle, we find we must finely tune the boundary conditions for $r\rightarrow
0$ for a solution to be reached. The reason is that any non-trivial solution
must obey ($r^{-2}\sin ^2F)$=1 to a very good accuracy for the range of
values $0<r<r_0\cong 1$ almost independently from the rest of the
differential equation, otherwise terms of the form ($r^{-2}\sin ^2F)^m$ go
to $0$ or $\infty $ for m large. Any account to parameterize the phase
transition must take that into account, which is not the case in our
procedure. We find that the $\alpha _1$, $\beta _2$ and $\lambda $
parameters must also be finely tuned, and that the slopes at $0$ and $\infty
$ do not always correspond to their numerical values.

Various ways of improving the fit of the solution may be taken into
consideration. For instance, we have tried using other trial functions, not
based on the Atiyah-Manton ansatz. The form of those is rather arbitrary
however and, despite somewhat improved results, we have still met the same
basic limitations. Another alternative we considered was to add a few more
terms to our series in order to account for the coefficients $\alpha _{11}$,
$\beta _{12}$, etc. Unfortunately, besides the rapidly increasing complexity
of our calculations, we have found that those terms contribute little to the
$r\sim \lambda $ region.

Since it seems impossible to carry out a direct fit of the solution in this
intermediate region, another approach we are considering is to modify the
corrective terms of the series, attempting to find a pattern in the values
taken by the $c_i$ coefficients that would make the series converge to a
closed form.

Even though imperfect, we stress again that this approach is still probably
the best alternative to complete numerical treatment, and that it can prove
very useful whenever an analytical form of the solution is required, as we
have already discussed with regard to stability analysis. The approach
Igarashi et al. followed with the Skyrme model can easily be generalized to
other cases (two-phase structure models, multi-Skyrmion configurations, ...)
by using a new family of trial functions based on our prototype.

\section{Acknowledgement}

One of the author would like to thank T.N. Truong and, the Centre de
Physique Th\'eorique where part of this work was done, for their
hospitality. This research was supported by the Natural Science and
Engineering Research Council of Canada and by the Fonds pour la Formation de
Chercheurs et l'Aide \`a la Recherche du Qu\'ebec.

\newpage



\begin{table}
\caption{The values of the $\alpha _1$ and $\beta _2$ coefficients (see
equations 13 and 15) and of some physical observables we have evaluated
with our analytical ansatz in the $N=1 $ case (single Skyrmion). The
Skyrmion mass $M_S$ (see equation 10) and the electrical radius
$<r^2>^{1/2}$ are expressed in units of  $\frac{F_\pi }{e } $  and $\frac{1
}{eF_\pi } $ respectively (with $e $ and $F_\pi $ defined in (2)), the pion
decay constant in MeV and the axial coupling constant $g_A$ is adimensional.
We also give the margin of error with respect to exact numerical results as
found in references [2-5].}
 \begin{tabular}{cdddddd}
    \centering{$\chi (a) $}  &  {$\alpha _1$}  &  {$\beta _2$}  &  {$M_S$ }
&  {$F_\pi $}  &  {$g_A$}  &  {$<r^2>^{1/2}$}  \\ \hline
   \centering{$\chi_S (a) $}      &  2.837       &  1.075       &  103.129
    &  64.59       &  -0.612       &  0.7479  \\
          &  (0.10\%)      &    (0.56\%)     &    (0.0004\%)     &  (0.03\%)
      &  (0.10\%)       &  (0.04\%)  \\
   \centering{$\chi_1 (a) $}      &  1.333     &  2.8062      & 136.505   &
72.984     & -0.713243       &  1.28496  \\
          &       &         &    (0.10\%)     &  (0.02\%)     &  (0.45\%)
   &  (0.50\%)  \\
   \centering{$\chi_{2,20} (a) $ }      &  0.981       &  2.050       &
73.5978       & 99.29       &  -0.743       &  1.2090  \\
          &         &        &  (3.59\%)       &  (3.86\%)       &  (20.9\%)
      &  (4.27\%)  \\
   \centering{$\chi_{3,20} (a) $}       &  0.970       &  2.440       &
98.3687       &  85.42       &  -0.766       &  1.2778  \\
          &  (1.25\%)       &          &  (0.89\%)       &  (0.75\%)       &
 (8.27\%)       &  (1.47\%)  \\
   \centering{$\chi_4 (a) $}      &  2.634       &  0.982       &  84.6796
    &  71.02      &  -0.697       & 0.7542  \\
          &         &        &  (0.01\%)       &         &         &
(0.03\%)  \\
   \centering{$\chi_5 (a) $}       &  2.587       &  1.034       &  88.2485
     &  70.54       &  -0.703       & 0.7784  \\
          &         &         &  (0.02\%)       &         &         &
(0.24\%)  \\
   \centering{$\chi_6 (a) $}       &  2.933       &  0.985       &  86.1798
     &  70.59       &  -0.697       &  0.7634  \\
          &         &        &  (0.40\%)       &         &         &
(0.07\%)  \\
   \centering{$\chi_7 (a) $ }      &  3.023       &  1.062       &  96.6275
     &  67.55       & -0.663       &  0.7824 \\
          &         &        &  (0.65\%)      &          &         &
(0.18\%)  \\
 \end{tabular}
\label{table1}
\end{table}

\begin{table}
\caption{The values of the $\alpha _1$ and $\beta _2$ coefficients (equation
13 and 15), of the Skyrmion mass $M_S$ (equation 10) and of the observables
$F_\pi $, $g_A$ and $<r^2>^{1/2}$ as obtained with our analytical ansatz in
the $N=2 $ case. The units are the same as in Table I . The margin of error
is given with respect to the numerical results we have obtained.}
 \begin{tabular}{cdddddd}
    \centering{$\chi (a) $}  &  {$\alpha _1$}  &  {$\beta _2$}  &  {$M_S$ }
&  {$F_\pi $}  &  {$g_A$}  &  {$<r^2>^{1/2}$}  \\\hline
   \centering{$\chi_S (a) $}      &  5.800       &  3.245       &  308.569
    &  40.94       &  -0.525       &  1.472  \\
          &  (4.30\%)      &  (0.37\%)       &  (0.48\%)      &          &
       &     \\
   \centering{$\chi_1 (a) $}      &  1.567     &  12.634      &  484.419
&        46.62  &    -0.555    &  2.993  \\
          &          &          &          &          &          &     \\
   \centering{$\chi_{2,20} (a) $ }      &  1.000       &  6.900       &
375.004       &  45.57       &  -0.717       &  2.5240  \\
          &          &          &          &          &          &     \\
   \centering{$\chi_{3,20} (a) $}       &  0.957       & 9.400       &
447.664       &  43.98       &  -0.692       &  2.8388 \\
          &          &          &          &          &          &     \\
   \centering{$\chi_4 (a) $}      &  5.360      &  2.934       & 265.890
 &  42.89       & -0.535       & 1.4205 \\
          &          &          &          &          &          &     \\
   \centering{$\chi_5 (a) $}       &  5.400       &  3.070       &  273.490
     &  42.78       &  -0.533       &  1.4515  \\
          &          &          &          &          &          &     \\
   \centering{$\chi_6 (a) $}       &  5.843       & 3.052       &  273.665
    &  42.93       & -0.546       &  1.5209  \\
          &          &          &          &          &          &     \\
   \centering{$\chi_7 (a) $}      &  6.111       &  3.063       &  294.382
    &  41.72       &  -0.548       &  1.5346  \\
          &          &          &          &          &          &     \\
 \end{tabular}
\label{table2}
\end{table}

\end{document}